
\magnification=1200
\baselineskip=24pt plus 4pt minus 4pt

\def\simle{\mathrel{
   \rlap{\raise 0.511ex \hbox{$<$}}{\lower 0.511ex \hbox{$\sim$}}}}
\def\simge{\mathrel{%
   \rlap{\raise 0.511ex \hbox{$>$}}{\lower 0.511ex \hbox{$\sim$}}}}

\null
\vskip 1.3in
\centerline{\bf ABSTRACT}
\vskip 0.7in
The one-band Hubbard model at half-filling on a truncated
tetrahedron (C$_{12}$), and on the C$_{60}$ molecule is studied.
Within the Hartree-Fock
approximation, we find a magnetic-like instability of the ``Fermi sea''
towards a spin-ordered phase for an intermediate value of the coupling
${\rm (U/t)_c}\approx 2.6$. The ordered phase presents a spin
arrangement similar to that of the classical Heisenberg model defined
on the same clusters. We  study the
linear excitations around the Hartree-Fock ground states using the
random-phase
approximation. On a finite cluster, we expect that these results signal
the presence of a $rapid$ $crossover$ between a paramagnetic region and
a regime where spin correlations are important.
The  relationship with the Heisenberg limit (large ${\rm U/t}$)
is discussed. Finally, we comment on implications of our results for
purely repulsive models of superconductivity in alkali-metal-doped fullerenes.
\vfill
\eject

The recent discovery$^1$ of superconductivity in alkali-metal-doped fullerenes,
K$_3$C$_{60}$ and Rb$_3$C$_{60}$, has led to
considerable interest in the electronic structure and superconductivity
mechanism of this new class of organic solids.
The remarkable structure of C$_{60}$ leads to one active p-orbital
per carbon atom, and it has been suggested that a one-band Hubbard model
will capture at least some of the physics of the $\pi$-electrons
of the fullerenes.$^{2,3}$ The theoretical analysis of this simplified model
can lead to an understanding of correlation effects that are beyond
the range of one-electron approximations. The same
model is currently widely used in the study of the cuprate high-T$_c$
superconductors. We have recently begun an investigation of the
Hubbard model on fullerene clusters, concentrating in this paper on the
Hartree-Fock and the Random Phase approximations. Results obtained
using the Quantum Monte Carlo method and the spin-wave approximation in the
Heisenberg limit were previously reported in Ref. 4.

In this paper we  study the one-band Hubbard
model at half-filling (one electron per site) for two particular clusters:  the
truncated tetrahedron C$_{12}$, and the truncated icosahedron C$_{60}$. Our
main result is the appearance of a magnetic-like instability beyond a critical
value of the Hubbard coupling ${\rm (U/t)_c \approx 2.6}$.
For larger values of U/t we observe that a generalized
spin-density wave state (SDW), i.e. a state with ``classical''
spin order on the cluster, has an energy lower than the Fermi sea  constructed
from the tight-binding eigenstates.
This new spin-density wave state  reduces, in the ${\rm U=\infty}$ limit, to
the
classical spin configuration which minimizes the  antiferromagnetic
Heisenberg Hamiltonian on the  cluster.$^{5,4}$ This configuration is highly
non-trivial as was shown in Ref.5. Physical consequences of this
instability are discussed at the end of the paper.

Certainly, for a $finite$ cluster we do not expect that the exact ground
state will
present a sharp singularity at some value of ${\rm U/t}$ (unless a level
crossing occurs). This is an artifact of our approximation.
There is no phase transition leading to magnetic ordering on a finite system
such as C$_{60}$. In the case of C$_{12}$ it has been shown$^3$ by exact
diagonalization of the Hubbard Hamiltonian that the ground state of
the neutral molecule is a singlet for all values of U/t. We also
expect this to be true in the case of  C$_{60}$. However it may very
well be that the physics of these {\it finite} systems is captured
by a broken symmetry state. For example,
this is known to happen in the case
of the square lattice Heisenberg antiferromagnet:
in the thermodynamic limit the quantum ground state is N\'eel ordered
(according to the general belief) and nevertheless the ground state
of any finite piece of square lattice is a singlet as demonstrated
long ago. But antiferromagnetic correlations indicative of N\'eel order
are present in these finite systems as clearly shown by numerical studies.$^6$
The  ground state degeneracy that we expect due to broken symmetry
appears only in the thermodynamic limit. For any finite system
vector order parameters will be zero (such as the staggered magnetization
of the square lattice magnet) but correlations can be properly studied
through the use of singlet correlation function such as $<{\bf S}_i \cdot
{\bf S}_j >$. Thus, we expect that the exact ground state of C$_{60}$ or
C$_{12}$
will show indications of the magnetic
instability discussed in this paper by presenting a $rapid$ $crossover$ at
${\rm (U/t)_c}$ from nonmagnetic to magnetic dominated correlations. The
ground state will always remain a singlet, but after the crossover a
state of
spin one will reduce its energy approaching the ground state when the
size of the clusters grow.

The Hamiltonian of the Hubbard model is
$$
 H= -{\rm t}\sum_{\langle i,j\rangle,\sigma}
( c_{i\sigma}^\dagger c_{j\sigma}^{
}+{\rm h.c.} ) +{{\rm U}\over 2}\sum_{i\sigma} c_{i\sigma}^\dagger  c_{i
\sigma}^{ } c_{i-\sigma}^\dagger c_{i-\sigma}^{ }  .
\eqno(1)
$$
The indexes $ i,j$ label sites of the cluster
($i,j$ = 1, ..., 60 for ${\rm C_{60}}$), $c_{i\sigma}^\dagger$  is
the creation operator for an electron on site $i$ with spin $\sigma$, t is the
hopping integral that we assume to be the same for all bonds, U is the
on-site repulsion energy, and the first sum runs over all neighbouring sites.
It is worth recalling that a simple tight-binding Hamiltonian for $\pi$
orbitals
reproduces extremely well the order and degeneracies of electronic levels near
the Fermi energy as obtained in more sophisticated molecular orbital
calculations.$^7$ At half-filling the only parameter of the model is the ratio
U/t. Its precise  value is unknown, but the intermediate regime
U/t$\approx$3-5 may
be realistic.$^{2,3,8}$ Recently, in the framework of a two-band Hubbard-like
model,
R. Lof {\it et al.}$^2$ claimed that doped ${\rm C_{60}}$
should be considered as a
highly correlated system with ${\rm U/W}$ comparable to that in high-${\rm
T_c}$ superconductors (${\rm W}$ being the one hole bandwidth). Then, it is
clearly interesting to study strongly correlated electrons defined on
fullerene clusters for $all$ values of ${\rm U/t}$.

The HF method searches for the Slater
determinant $|HF\rangle$ which minimizes the energy. This determinant is
built from the lowest eigenstates of the following one-particle Hamiltonian:
$$
h= -{\rm t}\sum_{\langle i,j\rangle,\sigma} ( c_{i\sigma}^\dagger
c_{j\sigma}^{ }+{\rm h.c.} ) +{\rm U}\sum_{i,\sigma}\left\{  n_{i-\sigma}\,
c_{i\sigma}^\dagger c_{i\sigma}^{ }
-\rho_{i\sigma,i-\sigma}\, c_{i\sigma}^\dagger
c_{i-\sigma}^{ }\right\} .
\eqno(2)
$$
In this equation,
$\rho_{i\sigma,j\sigma '}=\langle HF|c_{j\sigma'} ^\dagger
c_{i\sigma} ^{ }|HF\rangle$ is the one-body
density matrix, and $n_{i\sigma}\equiv\rho_{i\sigma,i\sigma}$.
One possible Ansatz for the density matrix, which always
provides a self-consistent solution, is $n_{i\uparrow}=n_{i\downarrow}=1/2$ and
$\rho_{i-\sigma,i\sigma}=0$. The interaction
term in $h$, Eq.(2),
is then proportional to the particle number operator and  the
eigenstates of $h$ are those of the tight-binding Hamiltonian. We refer to this
solution as to the trivial or ``non-magnetic'' state. Its energy  grows
linearly with U, starting from the ground state energy of the tight-binding
Hamiltonian. Its energy per site is, in units of t, -1.5+0.25(U/t) for
C$_{12}$ and -1.5527+0.25(U/t) for C$_{60}$. Of course,
the energy of this non-magnetic
state is overestimated by the Hartree-Fock approximation. As  U
increases, one expects the doubly occupied states to be depleted. The resulting
correlations could be taken into account by using a better variational state,
e.g. the Gutzwiller projection of this Slater determinant.$^8$ However,
when U becomes large enough, it is possible to find a better Slater
determinant which takes into account some of the correlations induced by the
on site repulsion. We turn now to the description of this new solution.

In the strong coupling regime (U/t $\gg$ 1) the half-filled
Hubbard model reduces at leading order in perturbation theory to the
antiferromagnetic Heisenberg model (HAF):
$$
H_{\scriptscriptstyle HAF}={4{\rm t}^2\over {\rm U}} \sum_{\langle i,j \rangle}
( {\bf S}_i \cdot
{\bf S}_j - {1 \over 4} ),
\eqno(3)
$$
where ${\bf S}_i$ are spin-$1/2$ operators. To study this quantum model
on finite clusters, one can search for the ground state of the corresponding
$classical$ model and perform the spin-wave expansion around this state.
This non-trivial ground state
classical spin configuration has been recently discussed.$^{5,4}$
For the quantum problem, a spin wave analysis
has been performed,$^4$ leading to a critical spin value (below
which spin order is washed out by quantum fluctuations) rather close to
1/2.
This result was in rough agreement with more sophisticated QMC simulations.$^4$
Here, we construct a
Hartree-Fock solution based on the classical spin configuration obtained by
minimizing the
Hamiltonian Eq.(3). For this purpose, we define a new basis $|i\sigma \rangle$
where the quantization axes are site-dependent, and lie in the directions of
the classical
spins. These directions are specified by polar angles
$\theta_i$ and $\phi_i$. In this new basis the Hamiltonian reads
$$
H=-{\rm t}\sum_{\langle i,j\rangle ,\sigma,\sigma'} ( (R_i^\dagger R_j)_{\sigma
\sigma'} c_{i\sigma}^\dagger c_{j\sigma'}^{ } +{\rm h.c.})
+{{\rm U}\over 2}\sum_{i,\sigma} c_{i\sigma}^\dagger c_{i\sigma}
^{ }c_{i-\sigma}^\dagger c_{i-\sigma}^{ },
\eqno(4)
$$
where $R_i$ is an $SU(2)$ matrix which rotates the original
quantization axis at site $i$
into the direction of polar angles $\theta_i,\phi_i$. Next, and after
the HF approximation is made in Eq.(4), we
try an Ansatz for which $\rho_{i\sigma,i-\sigma}=0$ and
$n_{i\uparrow}-n_{i\downarrow}$ is site-independent. This solution, when it
exists, will be called the ``magnetic''  solution. The corresponding  state
may be viewed as  a generalized
spin-density wave defined on clusters with the topology of a sphere.

The calculation of the classical ground state of the HAF has been
carried out for
C$_{12}$ and C$_{60}$. Using these results,
our HF calculations indicate that there is a magnetic solution beyond a
critical value (U/t)$_c \sim 2.6$ for C$_{60}$, and $\sim
2.7$ for C$_{12}$ for which
the order  parameter $n_{i\uparrow}-n_{i\downarrow}$ becomes $nonzero$.
For larger values of ${\rm U/t}$,
the magnetic solution is always lower in energy than the
nonmagnetic one. Figure 1 shows the value of $n_{i\uparrow}$ for C$_{60}$
as a function of U/t. The curve for C$_{12}$ has a similar shape.
In the limit of large U/t, $n_{i\uparrow}\to 1$, and thus
the spins align in the directions of the classical order
($\theta_i, \phi_i$). In the same limit, the HF energy, which behaves as
$t^2/U$, tends  towards the energy of the ground state of the equivalent
classical HAF. Figure 2 shows the energy per site for the C$_{60}$
cluster, in units of t, as a function  of U/t, for both the magnetic and the
non-magnetic solutions. The energies per site  of the classical ground
states are
$-2.5(t^2/U)$ for C$_{12}$, and $-2.809(t^2/U)$ for C$_{60}$, to be compared
against the energy per site of the
magnetic HF solution calculated for $(U/t)=5$ and $100$ which are,
respectively, $-2.395(t^2/U)$
and  $-2.499(t^2/U)$ for C$_{12}$, and $-2.64(t^2/U)$ and $-2.808(t^2/U)$ for
C$_{60}$. Finally, the gap between the highest occupied and the lowest
unoccupied orbitals  approaches $U$ for increasing values of $(U/t)$ since the
excitations in the Hartree-Fock approximation correspond to
flipping one spin against the mean field. Of course, the
interaction among these elementary spin flip excitations strongly alters this
picture, as revealed by the spin wave analysis or the RPA calculation to be
presented below. They clearly show that spin-wave-like states appear in the
HF gap.

The magnetic instability that we have found is qualitatively similar
to that observed for the Hubbard model on a lattice without nesting of
the Fermi surface, where
antiferromagnetism appears only above a critical (non vanishing) value of the
on-site repulsion.$^9$
This is to be contrasted with the case of the square lattice
where an antiferromagnetic instability  develops as soon as U is nonzero
due to perfect nesting of the Fermi surface: there, the non-magnetic solution
is
always unstable against the antiferromagnetic spin-density wave.

The HF magnetic solution breaks rotational symmetry. This
is an artifact of the mean-field approximation  and, of course, the true ground
state of a finite system, such as the C$_{60}$ cluster,
is an eigenstate of the total spin, presumably a singlet. However, the spin
correlations present in the Hartree-Fock state may be relevant to the physics
of the cluster. They may be found by the use of singlet observables.

Further insight on the spin correlations and the associated instability of the
non-magnetic Hartree-Fock solution may be obtained from the  Random Phase
Approximation (RPA). For C$_{12}$, we have performed  a complete RPA
calculation  in the space of the 144 particle-hole configurations,  for
both the magnetic and the  non-magnetic states. In the
case of C$_{60}$ we have only determined the lowest spin excited states from a
study of the spin susceptibility $
\chi_{ij}= -i\langle T S_i^z (t) S_j^z (0)\rangle ,
$
in real (lattice) space, and this only for
the non-magnetic solution. We find, in both cases of C$_{12}$ and C$_{60}$,
that the non-magnetic solution is stable against particle-hole excitations as
long as $(U/t)<(U/t)_c$. As $(U/t)$ crosses  the critical value, the lowest RPA
excitations energies, which are  degenerate (at least  3 times; for C$_{12}$
the total degeneracy is 9), do vanish and become imaginary. This
is illustrated in Fig.~3. This behaviour of the RPA eigenvalues reflects  the
fact that the non-magnetic HF solution is unstable for couplings
$(U/t)>(U/t)_c$.

We have checked explicitly, in the case of C$_{12}$, that the HF magnetic
solution is stable against small amplitude oscillations as given by the RPA.
Note that, in this case, the RPA spectrum contains  3 zero energy modes
reflecting the breaking of global spin rotational symmetry. We have
calculated, again for C$_{12}$,  the corrections to the total energy coming
from including RPA correlations in the ground state. In the strong coupling
limit, we find that the resulting energy  approaches,  as expected,$^{10,11}$
the linear spin-wave result for the limiting Heisenberg Hamiltonian, i.e.
$-3.313 (t^2/U)$  for C$_{12}$, and $-3.503 (t^2/U)$ for C$_{60}$.
The values obtained in RPA for U/t=5 and U/t=100 are
$-3.203(t^2/U)$ and $-3.3127(t^2/U)$, respectively.
Note that the RPA correlations, or
equivalently the spin waves, provide  a substantial correction to the
Hartree-Fock or classical energies. In the strong coupling limit, the 144 RPA
excitation energies of C$_{12}$ separate into a group of 12 lowest lying ones
whose values coincide with the  spin-wave energies of the equivalent HAF,
including the 3 zero energy states, and  another group of 132,
starting at an energy of order U, which  correspond to charge excitations.

Finally, the nature of the instability, as well as the structure of the non
trivial Hartree-Fock state, may be obtained from the RPA calculation in the
non-magnetic sector. To understand this, note that within RPA an
instability manifests itself  by the divergence of some static susceptibility.
In the present case, $\chi=\chi^0(1+{\rm U}\chi^0)
^{-1}$, where $\chi^0$ denotes the unperturbed spin
susceptibility, and the instability occurs  when  Det$(1+{\rm
U}\chi^0 (\omega =0) )=0$. The particular mode along which the instability
develops is found by the response of the system to a magnetizing field
${\bf h}_i$ pointing locally in the
directions $( \theta_i,\phi_i )$ of the
classical spins. The corresponding response function $\propto \chi_{ij}
{\bf h}_i \cdot {\bf h}_j$  diverges at U=U$_c$. There are indeed
three degenerate eigenvectors of the matrix $\delta_{ij}+{\rm U}\chi^0_{ij}
 (\omega =0)$ that
vanish right at U$_c$, whose components  can be identified with the
vectors
${\bf h}_i$. More precisely, if $X_i,Y_i,Z_i$ denote the components of
these three eigenvectors, we identify ${\rm h}_i^x=X_i,{\rm h}_i^y=Y_i,
{\rm h}_i^z=Z_i$, while the norm of the vector ${\bf h}_i$, i.e.
$X_i^2+Y_i^2+
Z_i^2$, is site independent. The relative angles between the orientations
of the magnetizing field  at sites i and j are then given by
$\cos\theta_{ij}=({\bf h}_i \cdot {\bf h}_j)/{\rm h}^2$, where h is
the strength of the magnetizing field, common to all sites. Thus, we recover
 precisely the configuration that is found in the Heisenberg limit and in the
HF calculation. We should emphasize that we have done this construction only
for the magnetic instability. The possibility of other
types of instabilities cannot be excluded. Actually, we observed
in the case of  C$_{12}$, for which a
complete RPA calculation has been performed, the presence of
nine RPA eigenvalues
which vanish as U approaches U$_c$. We still do not have a
simple physical interpretation to offer for these extra soft modes and
the possible associated instabilities.

To summarize, we have found and studied a generalized
spin-density wave state
for the Hubbard model on the clusters C$_{12}$ and C$_{60}$. This state is
lower in energy than the normal Fermi sea constructed from tight binding
orbitals when  U/t$\simge 2.6$. Qualitatively, it resembles the spin order
found by minimizing the energy of the classical Heisenberg model defined
on these clusters. We expect that our approximate broken symmetry solution
leads to a smooth crossover towards spin correlations on the finite system
as already advocated in ref. 5 on general grounds.
This magnetic instability may have an
important effect on the effective interaction between valence electrons. We
note in this respect that the perturbative calculations presented in Ref. 2
indicate that an attraction sets in for a value of the coupling U/t$\sim 3$.
It has been argued that such an attraction could be the origin of
superconductivity in doped C$_{60}$. Exact studies$^3$ of small
clusters, including C$_{12}$, have confirmed the existence
of an intermediate range of values of U/t where the pair-binding
energy is positive (equivalent to an effective attractive interaction
between electrons). It is interesting to note that this pairing
tendency appears in the range where spin correlations can be
expected to appear smoothly in the finite system.

We have checked that the new Hartree-Fock ground state
is stable against small fluctuations as calculated in the RPA. However, this
concerns only local stability, and we cannot  exclude the possible existence
of other minima in the Hartree-Fock energy. An important
question regarding the magnetic structure that we
have found is whether the spin correlations extend over the whole
molecule or if there is a finite spin correlation length smaller than the
molecular diameter. This requires further studies. Let us just
mention that for the C$_{12}$ cluster in the Heisenberg limit, we have
diagonalized the spin Hamiltonian in the $S^z =0$ sector and computed the
exact spin correlations $\langle {\bf S}_i \cdot {\bf S}_j \rangle$. For
the maximal separation (3 sites), the correlation is  reduced by a factor
$\simle $5 from the classical value but does not disappear. Thus it seems
plausible that the SDW state is not washed out by quantum fluctuations on
C$_{12}$. For the considerably larger C$_{60}$ molecules, the same question
has been addressed recently by {\it ab initio} techniques such as Quantum
Monte-Carlo. There is evidence$^4$ that the fluctuations are stronger in
C$_{60}$, leading perhaps to a spin liquid ground state$^{12}$ that resembles
the SDW here discussed only below the spin correlation length. However, the
QMC simulations were performed at finite temperature and, as recently remarked
in Ref.5, it may occur that reducing further the temperature in those
simulations,
an ordered state may be reached. This and related issues will be addressed
in the future.

The computer calculations were done at
the Cray-YMP of the Florida State University Computing Center,
Tallahassee, Florida. We thank this supercomputer center for its support.
\vfill
\eject
\centerline{\bf REFERENCES}
\bigskip
\item{[1]}A. Hebard et al., Nature (London) {\bf 350}, 600 (1991);
K. Holczer et al., Science {\bf 252}, 1154 (1991), see also H. Kroto et
al., Nature (London) {\bf 318}, 162 (1985).
\medskip
\item{[2]}S. Chakravarty and S. Kivelson, Europhys.
Lett. {\bf 16}, 751 (1991); S. Chakravarty, M. Gelfand and S. Kivelson,
Science {\bf 254}, 970 (1991); R. W. Lof et al., Phys. Rev. Lett. {\bf 68},
3924 (1992);
A. Auerbach and G. Murthy, BU preprints; W. Goff and P. Phillips, MIT
preprint. See also P. W. Anderson, talk at symposium
on ${\rm C_{60}}$ fullerenes, UCLA.
\medskip
\item{[3]}S. R. White, S. Chakravarty, M. P. Gelfand and S. Kivelson,
Phys. Rev. B{\bf 45}, 5062 (1992).
\medskip
\item{[4]}R. T. Scalettar, E. Dagotto, L. Bergomi, Th. Jolic\oe ur, H. Monien,
``Disordered Ground State of the Hubbard Model on a C$_{60}$ Cluster'',
FSU Preprint, SPhT/92-68.
\medskip
\item{[5]}D. Coffey and S. Trugman, LA-UR-91-3302 and 92-712, Los Alamos
preprints.
\item{[6]}E. Dagotto, Int. J. Mod. Phys. B{\bf 5}, 907 (1991) and references
therein.
\medskip
\item{[7]}D.A. Bochvar and E.G. Gal'pern, Dokl. Akad. Nauk. SSSR {\bf 209}, 610
(1973); S. Satpathy, Chem. Phys. Lett. {\bf 130}, 545 (1986).
\medskip
\item{[8]}P. Joyes and R. Tarento, Phys. Rev. B{\bf 45}, 12077 (1992).
\medskip
\item{[9]}D. Vollhardt, Rev. Mod. Phys. {\bf 56}, 99 (1984) and references
therein.
\medskip
\item{[10]}J. R. Schrieffer, X. G. Wen and S. C. Zhang, Phys. Rev. B{\bf 39},
11663 (1989).
\medskip
\item{[11]}A. V. Chubukov and D. M. Frenkel, University of Illinois preprint
P/92/34, and references therein.
\medskip
\item{[12]}L. Pauling, Proc. R. Soc. London {\bf A}196, 343 (1949); P. W.
Anderson, Science {\bf 235}, 1196 (1987); S. Kivelson, D. Rokhshar and J.
Sethna, Phys. Rev. B{\bf 35}, 8865 (1987).
\medskip
\vfill
\eject
\centerline{\bf FIGURE CAPTIONS:}
\vskip 1.0in
\noindent
{\bf Figure 1:} The value of $N_z\equiv n_{i\uparrow}$ as a function of U/t for
C$_{60}$. Below ${\rm (U/t)_c} \approx$2.6 the Hartree-Fock ground state is
nonmagnetic,
and $n_{i\uparrow}=n_{i\downarrow}=1/2$. Above the critical value, the
quantization axis points in
the directions of the classical spins, and  $n_{i\uparrow}$ approaches
its fully saturated value
$n_{i\uparrow}=1$, as ${\rm U \rightarrow} \infty$.
\bigskip
\noindent
{\bf Figure 2:} The energy per site in units of t for the HF solutions on
C$_{60}$.
The nonmagnetic solution (upper curve) rises linearly with U. Beyond (U/t)$_c$,
the magnetic solution (lower curve) has lower energy.
\bigskip
\noindent
{\bf Figure 3:} The energy, in units of the HF gap, of the three soft
RPA modes which vanish
at (U/t)$_c$, indicating the presence of the magnetic instability.
\vfill
\eject
\nopagenumbers
\tolerance=10000

\hfuzz=5pt
\baselineskip 12pt plus 2pt minus 2pt
\centerline{\bf GENERALIZED SPIN DENSITY WAVE STATE}
\centerline{\bf OF THE HUBBARD MODEL ON C$_{12}$ AND C$_{60}$ CLUSTERS.}
\vskip 36pt
\centerline{L. BERGOMI, J. P. BLAIZOT$^{*}$,
Th. JOLIC\OE UR\footnote{*}{C.N.R.S. Research Fellow}}
\vskip 12pt
\centerline{\it Service de Physique Th\'eorique\footnote{**}
{\rm Laboratoire de la Direction des Sciences de la Mati\`ere
 du Commissariat \`a l'Energie Atomique}}
\centerline{\it C.E.  Saclay}
\centerline{\it F--91191 Gif-sur-Yvette Cedex, France}
\vskip 12pt
\centerline{\it and}
\vskip 12pt
\centerline{ E. DAGOTTO}
\vskip 12pt
\centerline{\it Department of Physics,}
\centerline{\it Center for Materials Research and Technology,}
\centerline{\it and}
\centerline{\it Supercomputer Computations Research Institute,}
\centerline{\it Florida State University, Tallahassee, FL 32306, USA}
\vskip 0.5in
\centerline{Submitted to: {\it Physical Review B: Rapid Communications}}
\vskip 2.5in
\noindent  June 1992

\noindent PACS No: 74.70, 75.10.J, 75.30.D \hfill SPhT/92-083

\vfill
\eject
\bye